\documentclass[a4paper,11pt]{article}
\usepackage{pos}

\title{Strange partner of $T_{cc}^+$ from lattice QCD in $D^{(*)}D_s^{(*)}$ scattering}

\author*[a,b]{Tanishk Shrimal}
\author[c]{Sara Collins}
\author[f]{Priyajit Jana}
\author[a,b]{Padmanath M.}  
\author[d,e]{Sasa Prelovsek}

\affiliation[a]{The Institute of Mathematical Sciences, CIT Campus, Chennai, 600113, India}
\affiliation[b]{Homi Bhabha National Institute, Training School Complex, Anushaktinagar, Mumbai 400094, India}
\affiliation[c]{Institut für Theoretische Physik, Universität Regensburg, Regensburg, Germany}
\affiliation[d]{Faculty of Mathematics and Physics, University of Ljubljana, Ljubljana, Slovenia}
\affiliation[e]{Jozef Stefan Institute, Ljubljana, Slovenia}
\affiliation[f]{Department of Physics, University of Cyprus, Nicosia, Cyprus}

\emailAdd{tanishk.shrimal@imsc.res.in}
\emailAdd{sara.collins@ur.de}
\emailAdd{padmanath@imsc.res.in}
\emailAdd{sasa.prelovsek@ijs.si}

\abstract{
We report on our lattice QCD study of coupled $DD_s^*$–$D^*D_s$ scattering in the $J^{P}=1^{+}$ channel and elastic $DD_s$ scattering in the $J^{P}=0^{+}$ channel, aimed at investigating the possible existence of $cc\bar{u}\bar{s}$ tetraquarks near threshold. The calculation uses CLS ensembles with $m_{\pi}\sim280,\mathrm{MeV}$, lattice spacing $a\sim0.09,\mathrm{fm}$, and spatial extents $L/a=24,32$. Finite-volume spectra are obtained from a variational analysis of two-point correlation matrices constructed from two-meson operator bases using distillation. The $l=0$ partial-wave scattering amplitudes are determined from the lattice spectra in multiple moving frames using Lüscher's formalism as well as a finite-volume implementation of the Lippmann–Schwinger equation. In both channels we observe small but nonzero shifts relative to the noninteracting spectrum, indicating weak meson interactions. The extracted physically plausible $S$-wave amplitudes show no pole structures near threshold.
}

\FullConference{The 42nd International Symposium on Lattice Field Theory (LATTICE2025)\\
2-8 November 2025\\
Tata Institute of Fundamental Research, Mumbai, India\\
}

\begin{document}
\maketitle

\section{Introduction}

Over the past two decades, extensive theoretical and experimental efforts have focused on the spectroscopy of exotic hadrons, namely states whose quark content extends beyond conventional quark–antiquark mesons and three-quark baryons. Many of these states were predicted well before their experimental observation, and their discovery has stimulated significant progress in theoretical descriptions of multiquark systems.

Among the proposed exotic configurations, doubly heavy tetraquarks with flavor structure $QQ\bar u\bar d$ and quantum numbers $J^P=1^+$, where $Q$ denotes a heavy quark, have long been regarded as promising candidates that are stable under strong interactions. The observation of the doubly charmed tetraquark $T_{cc}^+$ by the LHCb Collaboration \cite{LHCb:2021vvq,LHCb:2021auc} provides compelling evidence that such configurations can exist as hadronic states and has renewed interest in understanding their internal structure and binding mechanisms.

Motivated by this observation, it is natural to investigate the strange partner of $T_{cc}^+$ with quark content $cc\bar u\bar s$. Theoretical studies of this system using quark models, QCD sum rules, and effective field theories generally indicate that it is less likely to form a deeply bound state, and may instead appear as a near-threshold resonance or virtual state. Since these predictions rely on model assumptions about the underlying dynamics, a first-principles investigation using lattice QCD is essential.

Lattice QCD provides a nonperturbative framework for studying hadronic interactions directly from QCD. Previous lattice calculations have mainly focused on the $bb\bar u\bar s$ system, where a bound state has been reported below the $BB_s^*$ threshold \cite{Francis:2016hui,Junnarkar:2018twb,Hudspith:2020tdf,Meinel:2022lzo,Alexandrou:2024iwi,Tripathy:2025vao}. In contrast, the $cc\bar u\bar s$ system has not yet been explored in detail at the level of scattering amplitudes. While a few studies have extracted finite-volume energy levels in this channel \cite{Junnarkar:2018twb,Cheung:2017tnt}, the corresponding infinite-volume amplitudes and pole structures remain unknown. Given that possible tetraquark states are expected to lie close to relevant two-meson thresholds, an amplitude analysis is necessary to distinguish between bound, resonant, and virtual-state scenarios.

In this contribution, I have presented our recent lattice QCD study of elastic $DD_s$ scattering in the scalar channel and coupled $DD_s^*$–$D^*D_s$ scattering in the axialvector channel \cite{tdjf-4l5v}. The analysis is based on variationally determined finite-volume spectra obtained from two lattice ensembles, with different spatial volumes, and multiple inertial frames. Infinite-volume amplitudes are extracted using the conventional Lüscher’s formalism as well as a finite-volume implementation of the Lippmann–Schwinger equation.

The remainder of this proceedings is organized as follows. Section~\ref{sec:SetupDetails} describes the lattice setup and the extraction of finite-volume spectra. The resulting spectra are discussed in Section~\ref{sec:FiniteVolume}. Section~\ref{sec:scatana} presents the scattering amplitude analysis, and the main results. Finally, we summarize our conclusions in Section~\ref{sec:conclusions}.

\section{Setup Details}
\label{sec:SetupDetails}

\paragraph{Ensembles:}
This calculation uses two CLS $N_f=2+1$ Wilson-clover ensembles at a single lattice spacing $a\simeq0.086$ fm \cite{Bruno:2014jqa,Bali:2016umi,Bruno:2016plf}, with $m_{\pi}\simeq280$ MeV and $m_K\simeq467$ MeV. The spin-averaged $1S$ charmonium mass, corresponding to the charm quark mass utilized, is slightly heavier than the physical value, the details of which can be found in Ref.~\cite{Prelovsek:2020eiw}.

\paragraph{Scattering channels:}
We consider channels relevant for scalar ($J^P=0^+$), axialvector ($1^+$), and tensor ($2^+$) quantum numbers. The relevant two-meson channels are $DD_s$, $DD_s^*$, $D^*D_s$, and $D^*D_s^*$. The $DD_s$ channel contributes to the scalar sector in $S$-wave, while the coupled $DD_s^*$ and $D^*D_s$ channels contribute to the axialvector sector. The $D^*D_s^*$ channel can contribute to all three quantum numbers but opens at significantly higher energies and is therefore omitted in the scalar and axialvector analyses.

\paragraph{Interpolating operators and correlation analysis:}
To extract the finite-volume spectrum we employ a basis of single- and two-meson interpolating operators. Single-meson masses are determined using local operators $\bar q_1\Gamma q_2$ with Dirac structures
\begin{equation}
J^P=0^-:\ \gamma_5,~\gamma_5\gamma_t,
\qquad
J^P=1^-:\ \gamma_i,~\gamma_i\gamma_t .
\end{equation}
Two-meson operators of the form $D^{(*)}D_s^{(*)}$ are constructed from the same Dirac structures and projected onto the relevant cubic-group irreducible representations using the methods of Refs.~\cite{Piemonte:2019cbi,Prelovsek:2020eiw,Padmanath:2022cvl},
\begin{align}
O^{D^{(*)}D^{(*)}_s}(\vec{P}) =
\sum_{k,(i),(j)} A_{k,(i),(j)}
D^{(*)}_{(i)}(\vec{p}_{1k})
D^{(*)}_{s(j)}(\vec{p}_{2k}),
\quad
\vec{P}=\vec{p}_{1k}+\vec{p}_{2k}.
\end{align}
Local diquark–antidiquark operators are not included, since previous studies found negligible impact on the low-lying spectrum \cite{Cheung:2017tnt,Prelovsek:2025vbr}.

Correlation functions are computed using the distillation framework \cite{Peardon:2009gh} with $N_v=75$ ($45$) Laplacian eigenvectors on the large (small) volume. The resulting correlation matrices are analyzed using the variational method by solving the generalized eigenvalue problem
\begin{equation}
\mathcal{C}(t)v^{(n)}(t)=\lambda^{(n)}(t)\mathcal{C}(t_0)v^{(n)}(t),
\end{equation}
from which the finite-volume energy levels are extracted. Statistical uncertainties are estimated using 1000 bootstrap samples.


\section{Finite-Volume Spectrum}
\label{sec:FiniteVolume}

\begin{figure*}[ht]
\centering
\includegraphics[width=0.7\linewidth]{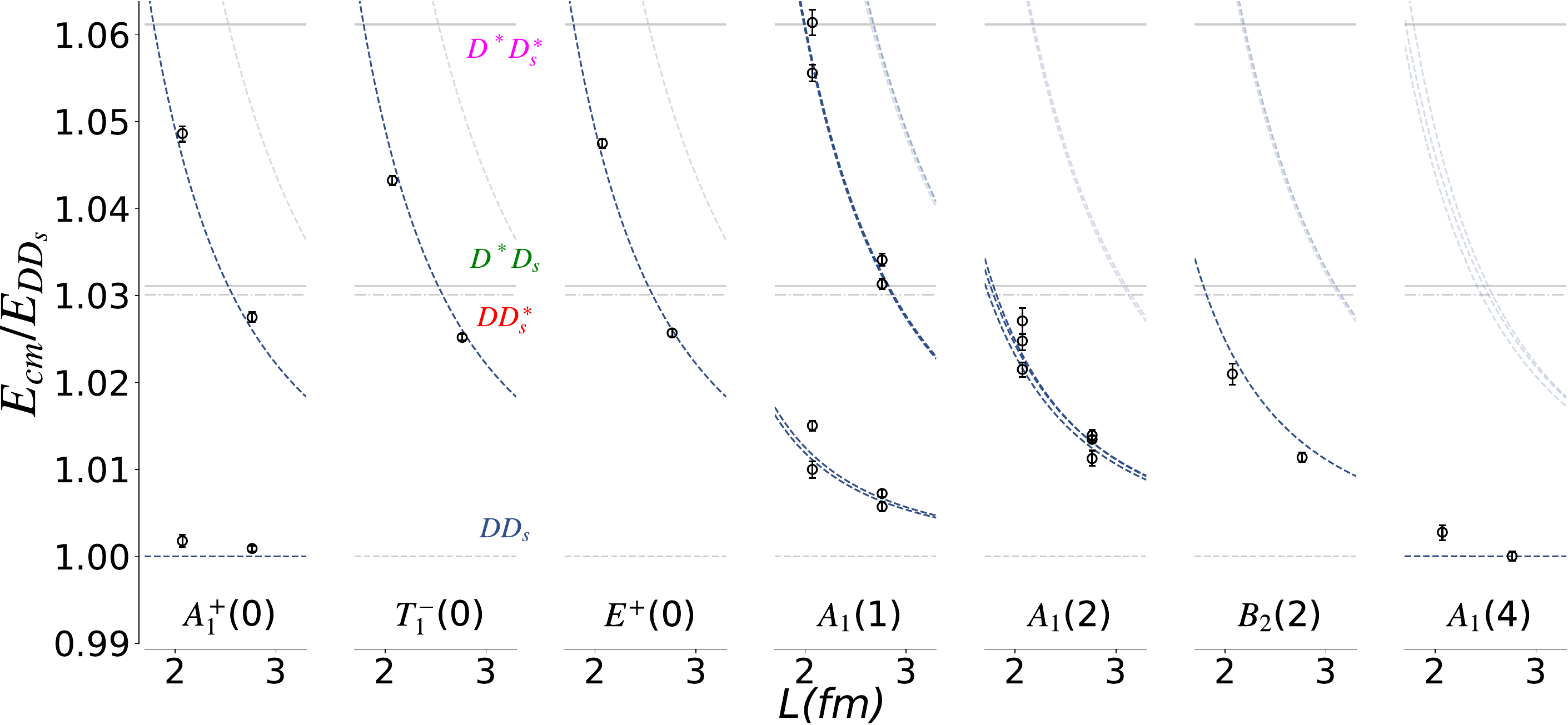}
\includegraphics[width=0.7\linewidth]{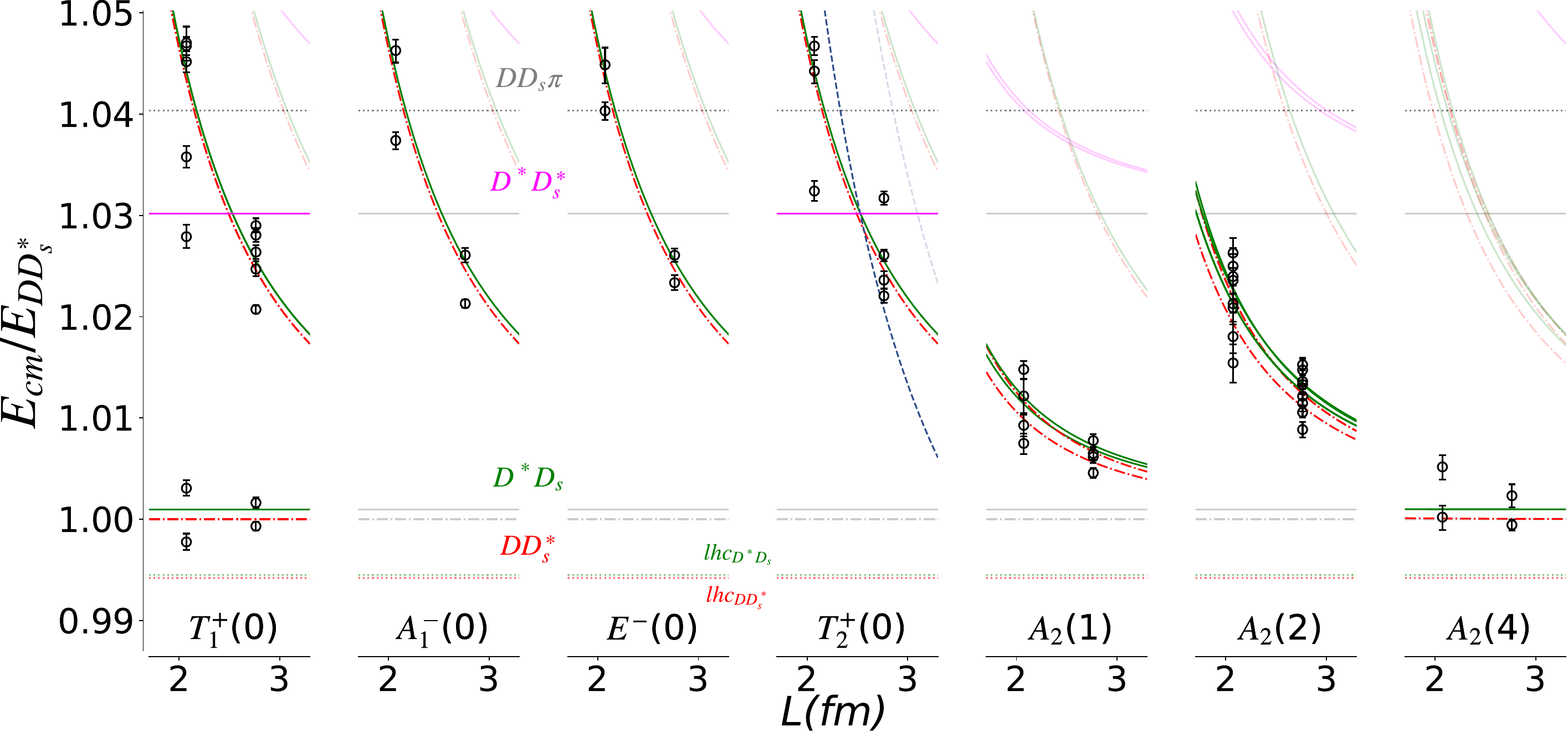}
\caption{Finite-volume energies in the center-of-momentum frame, expressed in units of the relevant thresholds and shown as a function of the spatial extent. The top (bottom) panel corresponds to irreps relevant for the scalar (axialvector) channel. Black markers denote lattice-extracted energies, while the curves indicate noninteracting two-meson levels for the channels $DD_s$ (blue dashed), $DD_s^*$ (red dot-dashed), $D^*D_s$ (green solid), and $D^*D_s^*$ (purple solid). Faded curves denote higher noninteracting shells not included in the analysis, and the gray dotted line indicates the $DD_s\pi$ threshold.}
\label{fig:full_spectrum}
\end{figure*}

Figure~\ref{fig:full_spectrum} shows the extracted finite-volume (FV) eigenenergies in the irreps relevant for the scalar and axialvector channels. Energies are presented in the center-of-momentum frame,
$E_{\rm cm}=\sqrt{E^2-\vec P^{\,2}}$, and compared with noninteracting two-meson levels constructed using the continuum dispersion relation with meson masses determined on the large volume. Deviations from these levels provide qualitative information about the underlying interactions.

\paragraph{Scalar channel ($J^P=0^+$):}
The lowest noninteracting levels are dominated by the $DD_s$ channel, allowing an elastic treatment near threshold. The ground-state energies in the relevant $S$-wave irreps show positive shifts relative to the corresponding noninteracting levels, indicating repulsive interactions between the $D$ and $D_s$ mesons. 

In contrast, irreps where the leading contribution arises from $P$-wave exhibit small negative shifts, suggesting weak attraction. Levels in irreps dominated by $D$-wave contributions remain consistent with noninteracting expectations within uncertainties. We therefore neglect $D$-wave interactions in the energy region considered.

\paragraph{Axialvector channel ($J^P=1^+$):}
In this sector the near-degenerate $DD_s^*$ and $D^*D_s$ thresholds require a coupled-channel treatment. The most relevant irrep is $T_1^+(0)$, where the lowest two levels show opposite shifts relative to the nearest noninteracting energies: the ground state is negatively shifted while the first excitation is positively shifted. This pattern is observed on both lattice volumes and becomes more pronounced on the smaller volume.

Overlap factors indicate that the lower level is dominated by $D(0)D_s^*(0)$ operators, while the higher level receives larger contributions from $D^*(0)D_s(0)$ operators, suggesting significant channel mixing.

\begin{figure}[t]
\centering
\includegraphics[width=0.46\linewidth]{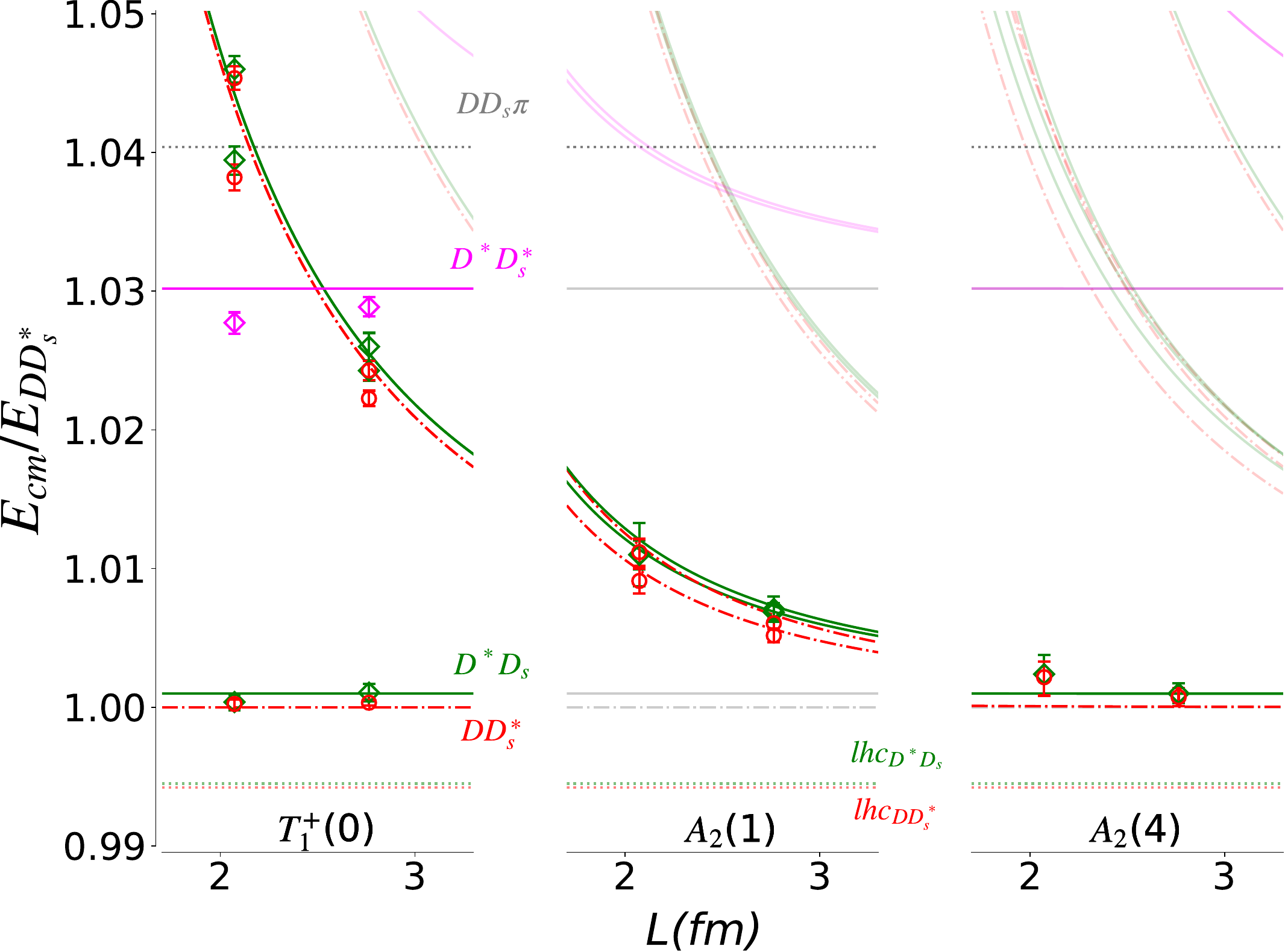}
\includegraphics[width=0.47\linewidth]{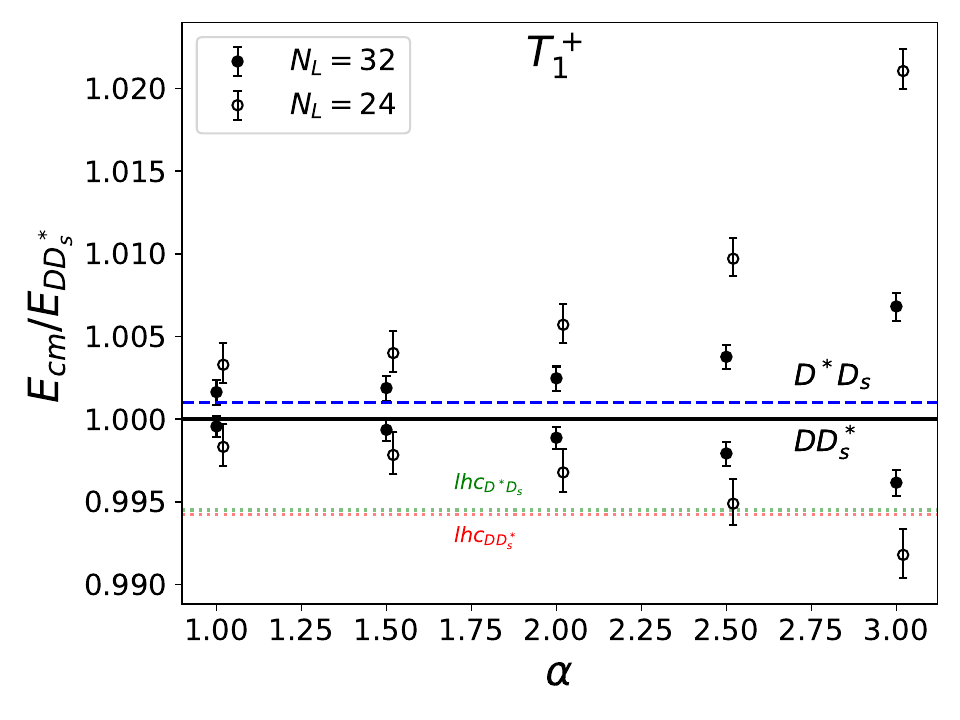}
\caption{Diagnostics of channel coupling in the $T_1^+(0)$ irrep.
\textit{Left:} spectra obtained from pruned operator bases excluding either $D^*D_s$ or $DD_s^*$ interpolators.
\textit{Right:} dependence of the lowest eigenenergies on the rescaling parameter $\alpha$ controlling cross-correlations between the two operator sectors.}
\label{fig:coupling_diagnostics}
\end{figure}

To assess the role of channel coupling, we perform two diagnostic tests shown in Fig.~\ref{fig:coupling_diagnostics}. First, we construct pruned operator bases by removing either $DD_s^*$-type or $D^*D_s$-type interpolators and analyze each channel separately. The resulting spectra are largely consistent with noninteracting expectations near threshold, indicating that the observed shifts primarily originate from coupled-channel dynamics.

Second, we artificially rescale the cross-correlations between the two operator sectors,
\begin{equation}
\tilde{\mathcal C}(t)=
\begin{pmatrix}
\mathcal C_{DD_s^*,DD_s^*}(t) &
\alpha\,\mathcal C_{DD_s^*,D^*D_s}(t) \\
\alpha\,\mathcal C_{D^*D_s,DD_s^*}(t) &
\mathcal C_{D^*D_s,D^*D_s}(t)
\end{pmatrix},
\end{equation}
where $\alpha=1$ corresponds to the simulated data. Increasing $\alpha$ enhances the splitting between the two lowest eigenvalues, confirming that the observed pattern originates from inter-channel coupling.

Levels dominated by $D^*D_s^*$-like operators appear close to the corresponding threshold and exhibit negative shifts, suggesting attractive interactions in this channel under an elastic assumption. However, the limited number of constraining levels prevents a quantitative amplitude determination.

Finally, contributions from higher partial waves are examined. While small shifts are observed in irreps dominated by $P$-wave, levels associated with $D$-wave operators remain consistent with the noninteracting scenario. We therefore neglect $D$-wave amplitudes in the subsequent scattering analysis.

\section{Scattering amplitudes and results}
\label{sec:scatana}

Following the extraction of the finite-volume spectra, the scattering amplitudes are determined using two complementary approaches.

\paragraph{Lüscher formalism:}
We employ the generalized Lüscher quantization condition
\begin{align}
\label{QCn}
\det\!\left[
(\tilde K^{(J)}_{cls;c'l's'}(E_{cm},\{a\}))^{-1}\delta_{JJ'}
-
\delta_{ss'}\delta_{cc'}
B^{\vec P,\Lambda}_{lJ;l'J'}(E_{cm})
\right]=0 .
\end{align}
Here $\tilde K$ is the reaction matrix, which is related to the infinite-volume scattering matrix up to known kinematic phase factors, while $B^{\vec P,\Lambda}$ is the finite-volume box matrix of known mathematical functions. The $\tilde K$-matrix parameters are determined by fitting the finite-volume spectrum to the lattice energies, allowing faithful extraction of the energy dependence of the infinite-volume scattering amplitude.
\paragraph{Lippmann--Schwinger approach:}
As an alternative approach, the amplitudes are obtained from a parametrized interaction potential by solving the Lippmann--Schwinger equation
\begin{align}
T(\vec{k},\vec{k}';E)
=
V(\vec{k},\vec{k}')
-
\int\!\frac{d^3q}{(2\pi)^3}
V(\vec{k},\vec{q})
G(\vec{q};E)
T(\vec{q},\vec{k}';E),
\end{align}
where $V$ denotes the interaction potential and $G$ the two-particle propagator. In finite volume the momentum integral becomes a discrete sum, leading to the determinant condition
\begin{equation}
\det[\mathbb{G}^{-1}(E)+\mathbb{V}(E)] = 0,
\end{equation}
whose solutions reproduce the finite-volume spectrum. Parameters of the $\tilde K$-matrix or interaction potential are determined from correlated $\chi^2$ fits to the lattice energies, and only parametrizations that describe the finite-volume spectrum satisfactorily are retained in the search for hadronic poles.

\subsection{Elastic $DD_s$ scattering in the scalar channel}

\begin{figure*}[ht]
\centering
\includegraphics[width=0.7\linewidth]{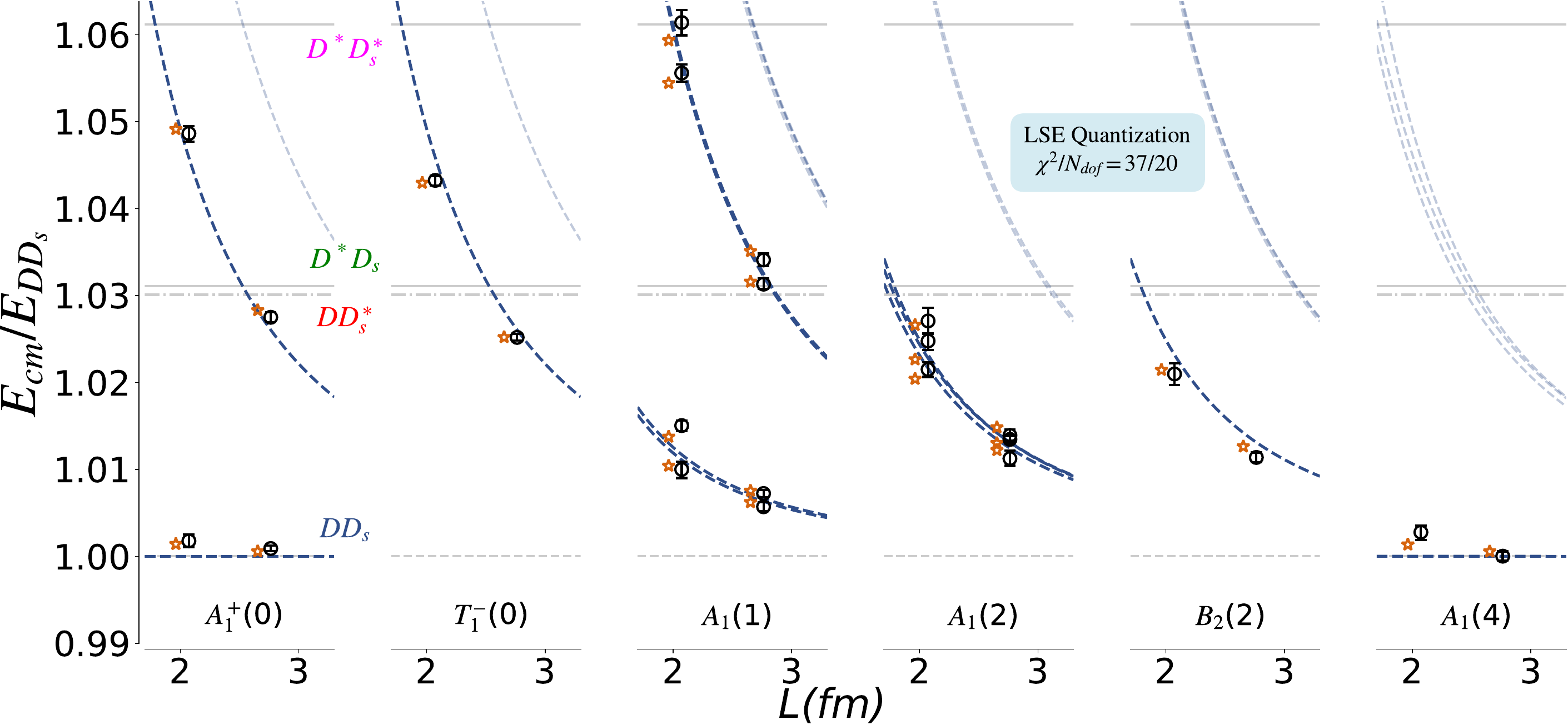}
\caption{Analytical reconstruction of the finite-volume spectrum using the best-fit parameters from combined $S$–$P$ wave fits obtained with the Lippmann--Schwinger approach. Orange stars denote reconstructed energies, while circles represent the lattice eigenenergies. Energies are expressed in units of the $DD_s$ threshold.}
\label{fig:scalar_chosen_spectrum}
\end{figure*}

\begin{figure}[ht]
\centering
\includegraphics[width=0.46\linewidth]{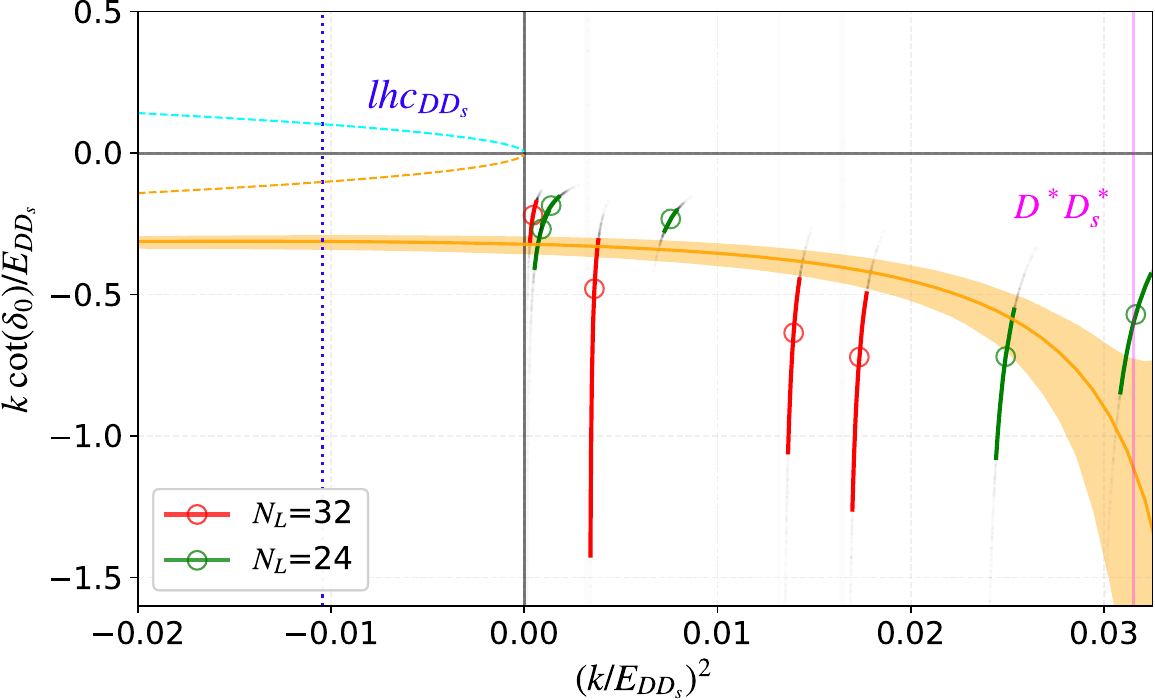}
\includegraphics[width=0.51\linewidth]{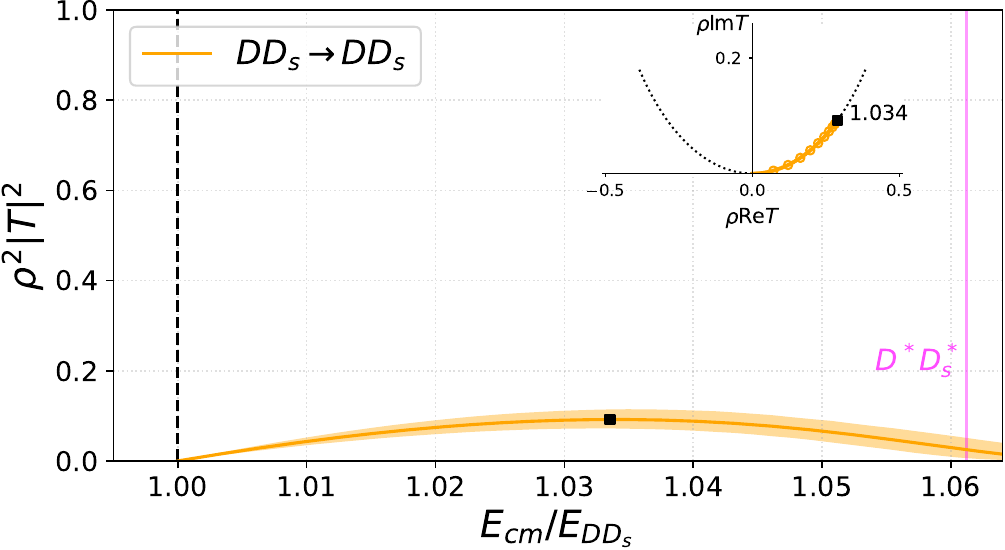}
\caption{left: $k\cot{\delta_0}/E_{DD_s}$ vs. $(k/E_{DD_s})^2$ for elastic $DD_s$ scattering in $S$-wave obtained from the LSE analysis. The markers correspond to lattice eigenenergies converted to $k\cot\delta$ using Lüscher's prescription. The orange curve and band represent the fitted amplitude. The cyan and orange curves denote the unitary parabola $\pm i\sqrt{-k^2}/E_{DD_s}$, while the vertical blue dotted line marks the left-hand-cut branch point from $K^*$ exchange. The magenta line indicates the $D^*D_s^*$ threshold. 
right: Energy dependence of the amplitude $\rho^2|T|^2$, proportional to the elastic cross section. The inset shows the Argand trajectory $\rho T=e^{i\delta}\sin\delta$. }
\label{fig:scalar_chosen_pcotdel}
\end{figure}

Our primary focus is the $S$-wave interaction in elastic $DD_s$ scattering. The finite-volume spectra are well described by both the Lüscher-based analysis and by solutions of the finite-volume Lippmann–Schwinger equation (LSE), each yielding acceptable $\chi^2/\mathrm{d.o.f.}$ values. Our preferred result is obtained from the LSE approach using a contact interaction that includes leading $S$- and $P$-wave contributions. The corresponding reconstructed finite-volume spectrum from the LSE analysis is shown in Fig.~\ref{fig:scalar_chosen_spectrum}. The result obtained using the Lüscher formalism is presented in the main article, with the associated scattering amplitude displayed in Fig.~\ref{fig:scalar_chosen_pcotdel}.

The resulting $S$-wave amplitude shows neither intersections with the bound-state constraint curves below threshold nor zero crossings above threshold. This indicates the absence of shallow bound states, virtual states, or resonances in the energy region constrained by the lattice data. The positive energy shifts of the finite-volume levels and the behavior of $k\cot\delta_0$ are consistent with a weakly repulsive $DD_s$ interaction.

The quantity $\rho^2|T|^2$ shows no pronounced peak, and the Argand trajectory exhibits only limited phase motion, further confirming the absence of resonant behavior. Lüscher-based parameterizations can generate subthreshold poles when unconstrained analytic extrapolations are used; such poles typically lie close to the left-hand cut or appear as acausal off-axis poles in the physical sheet and are therefore not assigned physical significance.

\paragraph{Higher partial waves.}

Although the focus is on $S$-wave scattering, moving-frame irreps require control of $P$-wave contributions. Fits to irreps where $P$-wave is the leading contribution show small but nonzero energy shifts, indicating weak interactions in this channel. Including $P$-wave terms in combined fits does not alter the extracted $S$-wave amplitude within uncertainties.

\subsection{Coupled-channel $DD_s^*$--$D^*D_s$ scattering}

\begin{figure*}[ht]
\centering
\includegraphics[width=0.7\linewidth]{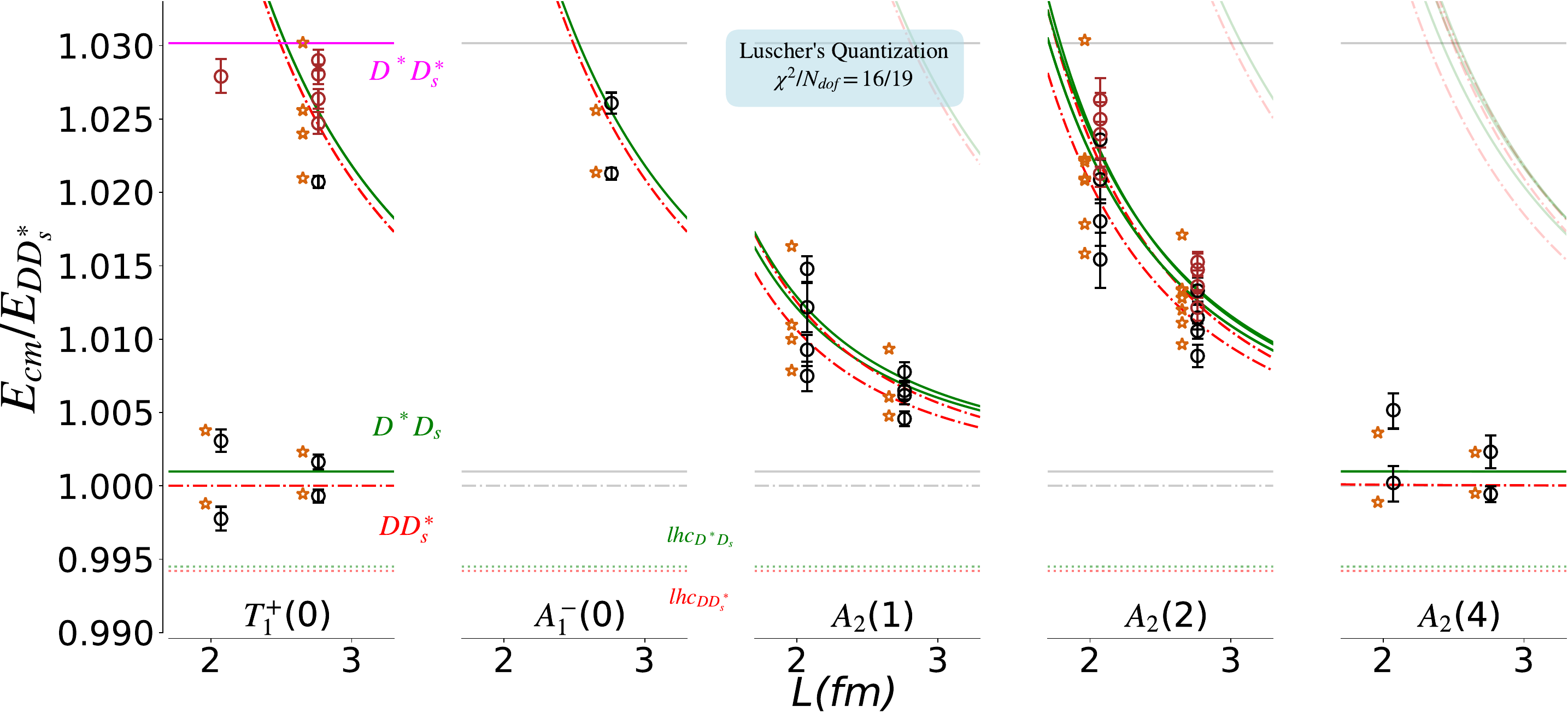}
\caption{Reconstructed spectra using the Lüscher's approach for the $DD_s^*$--$D^*D_s$ coupled-channel system in $S$-wave with $J^P=1^+$. Energies are shown in units of the $DD_s^*$ threshold. Red and green curves denote noninteracting levels. Black circles indicate levels used in the fits, while brown circles are excluded.}
\label{fig:avector_chosen_spectrum}
\end{figure*}

We now consider $S$-wave interactions in the coupled $DD_s^*$--$D^*D_s$ system with axialvector quantum numbers. Levels dominated by $l=2$ contributions are consistent with noninteracting expectations, so $D$-wave effects are neglected. Moving-frame irreps indicate nontrivial $P$-wave contributions, and we therefore perform both pure $S$-wave and combined $S$–$P$ wave analyses.

The reconstructed spectra are reasonably described by both LSE-based and Lüscher-based fits. Because of the limited number of energy levels and the larger parameter space in the coupled-channel system, the amplitudes are less tightly constrained than in the elastic case. Our preferred determination therefore uses Lüscher-based effective range expansion (ERE) parametrizations. The reconstructed spectrum from the Lüscher analysis is shown in Fig.~\ref{fig:avector_chosen_spectrum}, while the result from the LSE analysis is presented in the main article.
Using a two-channel $S$-wave ERE form we obtain scattering lengths
$a_{0,1} = 0.24^{+0.03}_{-0.02} \text{fm}$ and $a_{0,2} = -0.33^{+0.04}_{-0.02} \text{fm}.$

The off-diagonal interaction terms are significantly smaller than the diagonal components, indicating weak channel coupling. The amplitudes vary smoothly with energy, and the inelasticity remains close to unity. Neither channel exhibits zero crossings of $k\cot\delta$ above threshold nor intersections with the bound-state constraints below threshold, and a direct pole search reveals no singularities near threshold.

Alternative parametrizations, including Breit–Wigner forms, lead to consistent qualitative conclusions: weak interactions, small channel mixing, and no evidence for bound or resonant poles. Within the energy region constrained by the lattice data, the coupled $DD_s^*$--$D^*D_s$ system therefore shows no indication of a near-threshold tetraquark state.
\section{Conclusions}
\label{sec:conclusions}
In this exploratory study we have presented the first lattice QCD determination of the $D^{(*)}D^{(*)}_s$ scattering amplitudes in the scalar and axialvector channels at a single lattice spacing ($a\sim0.086$ fm) and $m_\pi \sim 280~\text{MeV}$. The finite-volume spectra were analyzed using both Lüscher’s formalism and a finite-volume implementation of the Lippmann–Schwinger equation.

For elastic $DD_s$ scattering in $S$-wave, the extracted amplitudes indicate a repulsive interaction and show no hadronic pole near threshold. Parameterizations lacking explicit analytic constraints can generate shallow bound or acausal off-axis poles, but such features are not supported by the more constrained Lippmann–Schwinger analysis. We also find that $P$-wave contributions do not significantly affect the extracted $S$-wave amplitude.

In the coupled $DD_s^*$–$D^*D_s$ system the near-threshold spectrum suggests weak interactions and small channel coupling. The extracted $S$-wave amplitudes exhibit smooth energy dependence and reveal no pole structures in the lattice-constrained region. Artificial enhancement of cross-correlations increases the apparent level splitting, suggesting some nontrivial channel-mixing effects in the $DD_s^*$–$D^*D_s$ system. Despite this, our results doesn't support the presence of a near-threshold $c\bar{c}\bar{u}\bar{s}$ bound or resonant state, in the kinematic region constrained by the lattice data.

The present study is performed at a single lattice spacing and heavier-than-physical light-quark masses, and does not include local diquark–antidiquark interpolators. Although previous studies suggest such operators have limited impact on low-lying spectra, systematic uncertainties associated with light-quark mass dependence, discretization effects (particularly in the charm sector), and finite-volume constraints remain to be quantified. Future calculations at lighter pion masses, multiple lattice spacings, and larger spatial volumes will be important to further constrain the amplitudes and clarify the existence of a strange partner of $T_{cc}^+$.


\begin{acknowledgments}
The authors acknowledge the support from the Slovenian Research Agency (research core Funding No. P1-0035), Department of Science and Technology, India (SERB Start-up Research Grant No. SRG/2023/001235), and Department of Atomic Energy, India. We thank our colleagues in CLS for the joint effort in the generation of the gauge field ensembles which form the basis for the computation. We use the multigrid solver of Refs.~\cite{Heybrock:2014iga,Heybrock:2015kpy,Richtmann:2016kcq,Georg:2017diz} for the inversion of the Dirac operator. Our code implementing distillation is written within the framework of the Chroma software package \cite{Edwards:2004sx}. The simulations have been performed on the Regensburg Athene2 cluster. We thank the authors of Ref. \cite{Morningstar:2017spu} for making the~{\it TwoHadronsInBox} package public.  We acknowledge the use of computing clusters at IMSc Chennai. We also thank the HPC RIVR consortium (www.hpc-rivr.si) and EuroHPC JU (eurohpc-ju.europa.eu) for funding this research by providing computing resources of the HPC system Vega at the Institute of Information Science (www.izum.si). 

\end{acknowledgments}

\bibliographystyle{JHEP} 
\bibliography{Tccus} 

@article{LHCb:2021vvq,
    author = "Aaij, Roel and others",
    collaboration = "LHCb",
    title = "{Observation of an exotic narrow doubly charmed tetraquark}",
    eprint = "2109.01038",
    archivePrefix = "arXiv",
    primaryClass = "hep-ex",
    reportNumber = "CERN-EP-2021-165, LHCb-PAPER-2021-031",
    doi = "10.1038/s41567-022-01614-y",
    journal = "Nature Phys.",
    volume = "18",
    number = "7",
    pages = "751--754",
    year = "2022"
}

@article{Bruno:2014jqa,
    author = "Bruno, Mattia and others",
    title = "{Simulation of QCD with N$_{f} =$ 2 $+$ 1 flavors of non-perturbatively improved Wilson fermions}",
    eprint = "1411.3982",
    archivePrefix = "arXiv",
    primaryClass = "hep-lat",
    reportNumber = "DESY-14-216, FTUAM-14-48, HIM-2014-01, HU-EP-14-51, MITP-14-091, SFB-CPP-14-89, IFT-UAM-CSIC-14-117",
    doi = "10.1007/JHEP02(2015)043",
    journal = "JHEP",
    volume = "02",
    pages = "043",
    year = "2015"
}

@article{Hudspith:2020tdf,
    author = "Hudspith, R. J. and Colquhoun, B. and Francis, A. and Lewis, R. and Maltman, K.",
    title = "{A lattice investigation of exotic tetraquark channels}",
    eprint = "2006.14294",
    archivePrefix = "arXiv",
    primaryClass = "hep-lat",
    doi = "10.1103/PhysRevD.102.114506",
    journal = "Phys. Rev. D",
    volume = "102",
    pages = "114506",
    year = "2020"
}

@article{Francis:2016hui,
    author = "Francis, Anthony and Hudspith, Renwick J. and Lewis, Randy and Maltman, Kim",
    title = "{Lattice Prediction for Deeply Bound Doubly Heavy Tetraquarks}",
    eprint = "1607.05214",
    archivePrefix = "arXiv",
    primaryClass = "hep-lat",
    doi = "10.1103/PhysRevLett.118.142001",
    journal = "Phys. Rev. Lett.",
    volume = "118",
    number = "14",
    pages = "142001",
    year = "2017"
}

@article{Junnarkar:2018twb,
    author = "Junnarkar, Parikshit and Mathur, Nilmani and Padmanath, M.",
    title = "{Study of doubly heavy tetraquarks in Lattice QCD}",
    eprint = "1810.12285",
    archivePrefix = "arXiv",
    primaryClass = "hep-lat",
    reportNumber = "TIFR/TH/18-43",
    doi = "10.1103/PhysRevD.99.034507",
    journal = "Phys. Rev. D",
    volume = "99",
    number = "3",
    pages = "034507",
    year = "2019"
}

@article{Alexandrou:2024iwi,
    author = "Alexandrou, Constantia and Finkenrath, Jacob and Leontiou, Theodoros and Meinel, Stefan and Pflaumer, Martin and Wagner, Marc",
    title = "{b\textasciimacron{}b\textasciimacron{}ud and b\textasciimacron{}b\textasciimacron{}us tetraquarks from lattice QCD using symmetric correlation matrices with both local and scattering interpolating operators}",
    eprint = "2404.03588",
    archivePrefix = "arXiv",
    primaryClass = "hep-lat",
    doi = "10.1103/PhysRevD.110.054510",
    journal = "Phys. Rev. D",
    volume = "110",
    number = "5",
    pages = "054510",
    year = "2024"
}

@article{Tripathy:2025vao,
    author = "Tripathy, Bhabani Sankar and Mathur, Nilmani and Padmanath, M.",
    title = "{$bb\bar u\bar d$ and $bs\bar u\bar d$ tetraquarks from lattice QCD using two-meson and diquark-antidiquark variational basis}",
    eprint = "2503.09760",
    archivePrefix = "arXiv",
    primaryClass = "hep-lat",
    journal = "Phys. Rev. D",
    reportNumber = "TIFR/TH/25-6",
    month = "3",
    year = "2025"
}

@article{LHCb:2021auc,
    author = "Aaij, Roel and others",
    collaboration = "LHCb",
    title = "{Study of the doubly charmed tetraquark $T_{cc}^{+}$}",
    eprint = "2109.01056",
    archivePrefix = "arXiv",
    primaryClass = "hep-ex",
    reportNumber = "CERN-EP-2021-169, LHCb-PAPER-2021-032",
    doi = "10.1038/s41467-022-30206-w",
    journal = "Nature Commun.",
    volume = "13",
    number = "1",
    pages = "3351",
    year = "2022"
}

@article{Bali:2016umi,
    author = {Bali, Gunnar S. and Scholz, Enno E. and Simeth, Jakob and S\"oldner, Wolfgang},
    collaboration = "RQCD",
    title = "{Lattice simulations with $N_f=2+1$ improved Wilson fermions at a fixed strange quark mass}",
    eprint = "1606.09039",
    archivePrefix = "arXiv",
    primaryClass = "hep-lat",
    doi = "10.1103/PhysRevD.94.074501",
    journal = "Phys. Rev. D",
    volume = "94",
    number = "7",
    pages = "074501",
    year = "2016"
}

@article{Prelovsek:2025vbr,
    author = "Prelovsek, S. and Ortiz-Pacheco, E. and Collins, S. and Leskovec, L. and Padmanath, M. and Vujmilovic, I.",
    title = "{Doubly heavy tetraquarks from lattice QCD: incorporating diquark-antidiquark operators and the left-hand cut}",
    eprint = "2504.03473",
    archivePrefix = "arXiv",
    primaryClass = "hep-lat",
    journal = "Phys. Rev. D",
    month = "4",
    year = "2025"
}

@article{Bruno:2016plf,
    author = "Bruno, Mattia and Korzec, Tomasz and Schaefer, Stefan",
    title = "{Setting the scale for the CLS $2 + 1$ flavor ensembles}",
    eprint = "1608.08900",
    archivePrefix = "arXiv",
    primaryClass = "hep-lat",
    reportNumber = "DESY-16-162, WUB-16-05",
    doi = "10.1103/PhysRevD.95.074504",
    journal = "Phys. Rev. D",
    volume = "95",
    number = "7",
    pages = "074504",
    year = "2017"
}

@article{Prelovsek:2020eiw,
    author = "Prelovsek, Sasa and Collins, Sara and Mohler, Daniel and Padmanath, M. and Piemonte, Stefano",
    title = "{Charmonium-like resonances with J$^{PC}$ = 0$^{++}$, 2$^{++}$ in coupled $ \mathrm{D}\overline{\mathrm{D}} $, $ {\mathrm{D}}_{\mathrm{s}}{\overline{\mathrm{D}}}_{\mathrm{s}} $ scattering on the lattice}",
    eprint = "2011.02542",
    archivePrefix = "arXiv",
    primaryClass = "hep-lat",
    reportNumber = "MITP/20-065",
    doi = "10.1007/JHEP06(2021)035",
    journal = "JHEP",
    volume = "06",
    pages = "035",
    year = "2021"
}

@article{Padmanath:2022cvl,
    author = "Padmanath, M. and Prelovsek, S.",
    title = "{Signature of a Doubly Charm Tetraquark Pole in DD* Scattering on the Lattice}",
    eprint = "2202.10110",
    archivePrefix = "arXiv",
    primaryClass = "hep-lat",
    reportNumber = "MITP/22-018",
    doi = "10.1103/PhysRevLett.129.032002",
    journal = "Phys. Rev. Lett.",
    volume = "129",
    number = "3",
    pages = "032002",
    year = "2022"
}

@article{Peardon:2009gh,
    author = "Peardon, Michael and Bulava, John and Foley, Justin and Morningstar, Colin and Dudek, Jozef and Edwards, Robert G. and Joo, Balint and Lin, Huey-Wen and Richards, David G. and Juge, Keisuke Jimmy",
    collaboration = "Hadron Spectrum",
    title = "{A Novel quark-field creation operator construction for hadronic physics in lattice QCD}",
    eprint = "0905.2160",
    archivePrefix = "arXiv",
    primaryClass = "hep-lat",
    reportNumber = "JLAB-THY-09-985",
    doi = "10.1103/PhysRevD.80.054506",
    journal = "Phys. Rev. D",
    volume = "80",
    pages = "054506",
    year = "2009"
}

@article{Piemonte:2019cbi,
    author = "Piemonte, Stefano and Collins, Sara and Mohler, Daniel and Padmanath, M. and Prelovsek, Sasa",
    title = "{Charmonium resonances with $J^{PC}=1^{--}$ and $3^{--}$ from $\bar DD$ scattering on the lattice}",
    eprint = "1905.03506",
    archivePrefix = "arXiv",
    primaryClass = "hep-lat",
    reportNumber = "MITP/19-029",
    doi = "10.1103/PhysRevD.100.074505",
    journal = "Phys. Rev. D",
    volume = "100",
    number = "7",
    pages = "074505",
    year = "2019"
}

@article{Morningstar:2017spu,
    author = {Morningstar, Colin and Bulava, John and Singha, Bijit and Brett, Ruair\'\i{} and Fallica, Jacob and Hanlon, Andrew and H\"orz, Ben},
    title = "{Estimating the two-particle $K$-matrix for multiple partial waves and decay channels from finite-volume energies}",
    eprint = "1707.05817",
    archivePrefix = "arXiv",
    primaryClass = "hep-lat",
    doi = "10.1016/j.nuclphysb.2017.09.014",
    journal = "Nucl. Phys. B",
    volume = "924",
    pages = "477--507",
    year = "2017"
}

@article{Meinel:2022lzo,
    author = "Meinel, Stefan and Pflaumer, Martin and Wagner, Marc",
    title = "{Search for b\textasciimacron{}b\textasciimacron{}us and b\textasciimacron{}c\textasciimacron{}ud tetraquark bound states using lattice QCD}",
    eprint = "2205.13982",
    archivePrefix = "arXiv",
    primaryClass = "hep-lat",
    doi = "10.1103/PhysRevD.106.034507",
    journal = "Phys. Rev. D",
    volume = "106",
    number = "3",
    pages = "034507",
    year = "2022"
}

@article{Cheung:2017tnt,
    author = "Cheung, Gavin K. C. and Thomas, Christopher E. and Dudek, Jozef J. and Edwards, Robert G.",
    collaboration = "Hadron Spectrum",
    title = "{Tetraquark operators in lattice QCD and exotic flavour states in the charm sector}",
    eprint = "1709.01417",
    archivePrefix = "arXiv",
    primaryClass = "hep-lat",
    reportNumber = "DAMTP-2017-33, JLAB-THY-17-2541",
    doi = "10.1007/JHEP11(2017)033",
    journal = "JHEP",
    volume = "11",
    pages = "033",
    year = "2017"
}

@inproceedings{Heybrock:2014iga,
    author = "Heybrock, Simon and Jo\'o, B\'alint and Kalamkar, Dhiraj D. and Smelyanskiy, Mikhail and Vaidyanathan, Karthikeyan and Wettig, Tilo and Dubey, Pradeep",
    title = "{Lattice QCD with domain decomposition on Intel\textregistered{} Xeon Phi\texttrademark{} co-processors}",
    booktitle = "{The International Conference for High Performance Computing, Networking, Storage, and Analysis}: {SC14: HPC matters}",
    eprint = "1412.2629",
    archivePrefix = "arXiv",
    primaryClass = "hep-lat",
    reportNumber = "JLAB-IT-14-03",
    doi = "10.1109/SC.2014.11",
    month = "12",
    year = "2014"
}

@article{Heybrock:2015kpy,
    author = "Heybrock, Simon and Rottmann, Matthias and Georg, Peter and Wettig, Tilo",
    title = "{Adaptive algebraic multigrid on SIMD architectures}",
    eprint = "1512.04506",
    archivePrefix = "arXiv",
    primaryClass = "physics.comp-ph",
    doi = "10.22323/1.251.0036",
    journal = "PoS",
    volume = "LATTICE2015",
    pages = "036",
    year = "2016"
}

@article{Richtmann:2016kcq,
    author = "Richtmann, Daniel and Heybrock, Simon and Wettig, Tilo",
    title = "{Multiple right-hand-side setup for the DD-$\alpha$AMG}",
    eprint = "1601.03184",
    archivePrefix = "arXiv",
    primaryClass = "hep-lat",
    doi = "10.22323/1.251.0035",
    journal = "PoS",
    volume = "LATTICE2015",
    pages = "035",
    year = "2016"
}

@article{Georg:2017diz,
    author = "Georg, Peter and Richtmann, Daniel and Wettig, Tilo",
    title = "{pMR: A high-performance communication library}",
    eprint = "1701.08521",
    archivePrefix = "arXiv",
    primaryClass = "hep-lat",
    doi = "10.22323/1.256.0361",
    journal = "PoS",
    volume = "LATTICE2016",
    pages = "361",
    year = "2017"
}

@article{Edwards:2004sx,
    author = "Edwards, Robert G. and Joo, Balint",
    editor = "Bodwin, Geoffrey T. and Sinclair, D. K. and Eichten, E. and Holmgren, D. and Kronfeld, Andreas S. and Mackenzie, P. and Okamoto, M. and Simone, J. and El-Khadra, Aida X.",
    collaboration = "SciDAC, LHPC, UKQCD",
    title = "{The Chroma software system for lattice QCD}",
    eprint = "hep-lat/0409003",
    archivePrefix = "arXiv",
    reportNumber = "JLAB-THY-04-54",
    doi = "10.1016/j.nuclphysbps.2004.11.254",
    journal = "Nucl. Phys. B Proc. Suppl.",
    volume = "140",
    pages = "832",
    year = "2005"
}

@article{tdjf-4l5v,
  title = {Lattice study of $cc\overline{u}\overline{s}$ tetraquark channels in ${D}^{(*)}{D}_{s}^{(*)}$ scattering},
  author = {Shrimal, Tanishk and Collins, Sara and Jana, Priyajit and Padmanath, M. and Prelovsek, Sasa},
  journal = {Phys. Rev. D},
  volume = {112},
  issue = {5},
  pages = {054513},
  numpages = {31},
  year = {2025},
  month = {Sep},
  publisher = {American Physical Society},
  doi = {10.1103/tdjf-4l5v},
  url = {https://link.aps.org/doi/10.1103/tdjf-4l5v}
}

\end{document}